\documentclass[reprint,twocolumn, superscriptaddress, floatfix,
 amsmath,amssymb,
 aps,
 longbibliography,
 footinbib,
]{revtex4-2}

\usepackage{graphicx}% Include figure files
\usepackage{dcolumn}% Align table columns on decimal point
\usepackage{bm}% bold math
\usepackage[dvipsnames]{xcolor}
\usepackage{comment}
\usepackage{physics}
\usepackage[colorlinks=true,bookmarks=false,citecolor=blue,urlcolor=blue]{hyperref}

\begin{document}

\title{\Large{Bound States in the Continuum in a Wire Medium}}
%\title{\Large{Bound States in the Continuum induced by Spatial Dispersion}}

\author{\firstname{E.}\,\surname{Koreshin}}
\affiliation{Qingdao Innovation and Development Base of Harbin Engineering University, Qingdao 266000, China}
\affiliation{School of Physics and Engineering, ITMO University, St. Petersburg 197101, Russia}
\author{\firstname{S.}\,\surname{Gladyshev}}
\affiliation{School of Physics and Engineering, ITMO University, St. Petersburg 197101, Russia}
\author{\firstname{I.}\,\surname{Matchenya}}
\affiliation{School of Physics and Engineering, ITMO University, St. Petersburg 197101, Russia}
\affiliation{Skolkovo Institute of Science and Technology, Moscow 121205, Russia}
\author{\firstname{R.}\,\surname{Balafendiev}}
\affiliation{School of Physics and Engineering, ITMO University, St. Petersburg 197101, Russia}
\author{\firstname{I.}\,\surname{Terekhov}}
\affiliation{School of Physics and Engineering, ITMO University, St. Petersburg 197101, Russia}
\author{\firstname{P.}\,\surname{Belov}}
\affiliation{School of Physics and Engineering, ITMO University, St. Petersburg 197101, Russia}
\author{\firstname{A.}\,\surname{Bogdanov}}
\email{a.bogdanov@metalab.ifmo.ru}
\affiliation{Qingdao Innovation and Development Base of Harbin Engineering University, Qingdao 266000, China}
\affiliation{School of Physics and Engineering, ITMO University, St. Petersburg 197101, Russia}
\date{\today}

\begin{abstract}
We show that a slab of wire medium composed of thin parallel metallic wires can naturally support bound states in the continuum (BICs) formed in an unusual way. The revealed BICs appear due to the strong spatial dispersion making possible the propagation of longitudinal plasma-like waves and TEM polarized modes with a flat band. The symmetry-protected (at-$\Gamma$) BICs are formed due to the polarization mismatch between the longitudinal plasma-like waves and transversal plane waves in the surrounding space, while the accidental (off-$\Gamma$)  BICs appear as a result of the destructive interference between bulk TEM and plasma modes. All revealed BICs can be well-described analytically without the use of the Bloch theorem within effective medium approximation when the wire medium behaves as homogeneous 1D anisotropic plasma with strong spatial dispersion.

\end{abstract}

\maketitle

{\it Bound states in the continuum} (BICs) are non-radiating resonances of open systems with a spectrum embedded in the continuum of the propagating waves of the surrounding space~\cite{Hsu2016Jul,Koshelev2023May,Liu2024Jun,Kang2023Nov}. BICs have attracted  much attention in recent years in photonics because of their pronounced resonant properties, scalability to any frequency range, and compatibility with various photonic structures and materials~\cite{Azzam2021Jan,Xu2023Jun,Kang2023Nov}. BIC possesses a diverging radiative quality factor (Q-factor), strong field localization, and topological robustness with respect to the variation system's parameters preserving its symmetry~\cite{Zhen2014Dec}. All these advantages make BICs very prospective for micro and nanolasers~\cite{Kodigala2017Jan,Yu2021Oct,Hwang2021Jul,Wu2020Aug}, biosensors~\cite{Romano2019Jun,Romano2018Jul}, polaritonics~\cite{Kravtsov2020Apr, Ardizzone2022May}, and nonlinear photonic components~\cite{Koshelev2019Jul,Liu2019Dec,Koshelev2020Jan}.

%At the same time, BICs can be interpreted as polarization vertices with a certain topological charge. It is robust against variations of geometrical parameters preserving the symmetry of the structure.

The BICs are mostly studied in periodic photonic structures, including both all-dielectric, plasmonic, spoof-plasmonic, or even made of perfect electric conducting materials ~\cite{li2022quasi,zhou2022plasmonic,ulrich1975modes,huang2021bound,Azzam2018,Liang2020,Li2022}  ones, where their origin is well-studied and understood~\cite{Azzam2018Dec,Koshelev2023May}. Periodicity makes the radiation possible only into the open {\it diffraction channels} that substantially simplifies observation BIC in practice, especially in structures with a subwavelength period where only two diffraction channels are open~\cite{Zhen2014Dec,Hsu2013Jul,Sadrieva2017Apr}. Another way to engineer BIC is the Friedrich-Wintgen scenario 
%Eu 
when the destructive interference of two resonant (leaky) states results in a complete cancellation of the radiation losses~\cite{Friedrich1985Dec}. Such behavior is usually associated with the strong coupling of two optical modes with a characteristic avoid crossing~\cite{bogdanov_bound_2019}. This mechanism can result in the appearance of BICs even in structures with continuous translation symmetry when destructive interference of the ordinary and extraordinary modes in planar anisotropic dielectric structures can result in complete suppression of radiation losses~\cite{Gomis-Bresco2017Apr}. In acoustic structures, the same effect is achieved due to the interference between shear and longitudinal waves~\cite{Quotane2018Jan,Mizuno2019Feb,Deriy2022Feb}.

In this study, we unveil a novel scenario for BIC formation in photonic structures exhibiting strong {\it spatial dispersion}. This spatial dispersion enables the propagation of additional waves that can give rise to BICs.  We illustrate this with the example of a {\it wire medium} composed of thin metallic wires, which can be effectively described as a homogeneous highly non-local anisotropic medium~\cite{Belov2003Mar}. Specifically, we demonstrate that both {\it symmetry-protected} and {\it accidental} BICs are formed in an extraordinary manner due to the strong spatial dispersion resulting in the appearance of plasma-like waves~\cite{Belov2003Mar}. They can solely form at-$\Gamma$ BICs or for off-$\Gamma$ BICs due to destructive interference with TEM modes. All these BICs can be accurately described within the framework of the effective medium approximation (EMA), whereas BICs in dielectric and plasmonic metasurfaces require consideration of the periodic potential using the Bloch theorem and the concept of diffraction channels.

\begin{figure}[t]
\centering
\includegraphics[width=0.99\linewidth]{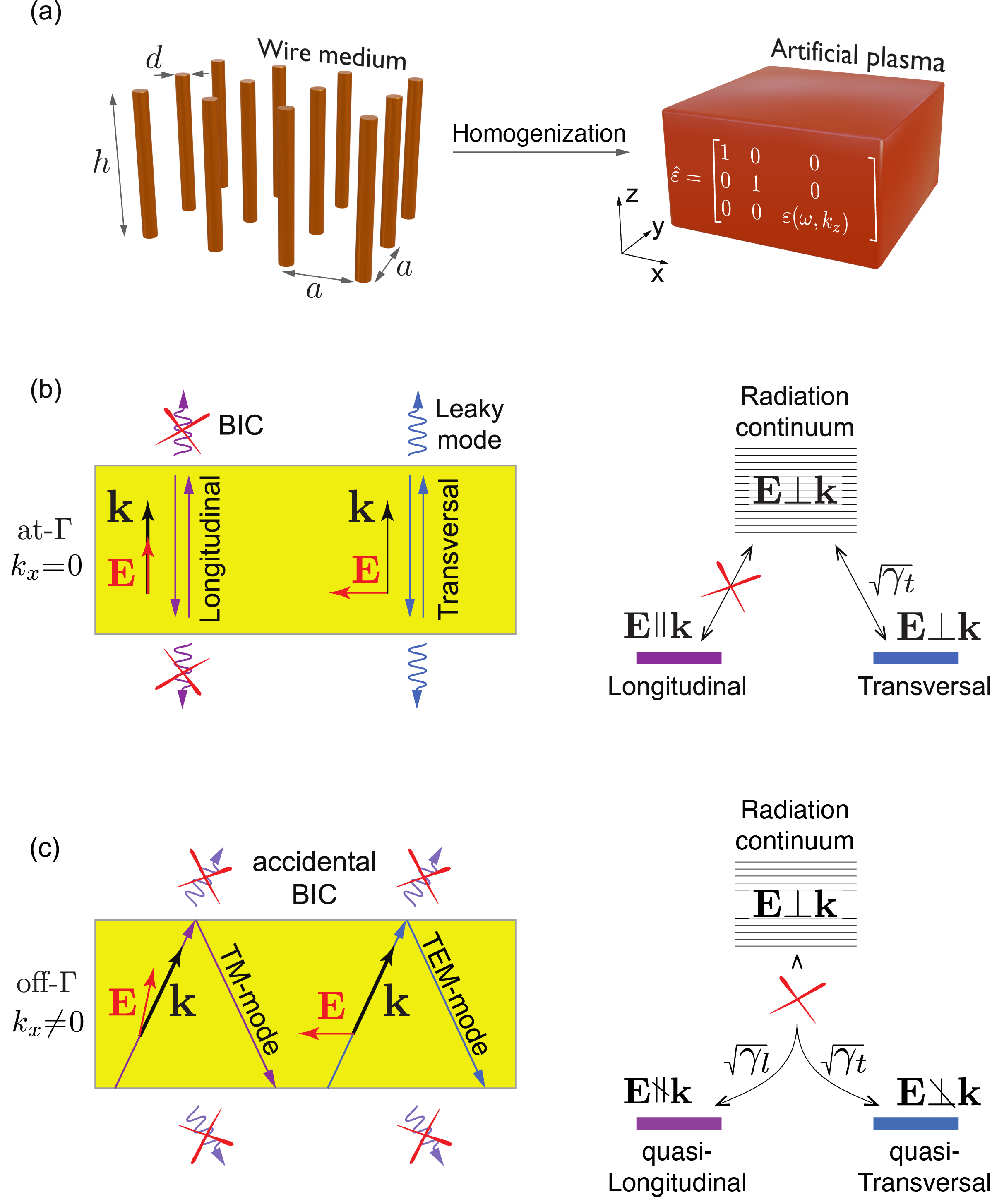}
\caption{(a) Slab of wire medium made of a perfect electric conductor. Within EMA, the slab can be described as anisotropic plasma with strong spatial dispersion [see Eqs.~\eqref{Eq:eq-matrix},\eqref{Eq:eps}, and \eqref{Eq:plasma-freq}]. (b) Modes of the slab at the $\Gamma$-point ($k_x=k_y=0$). Longitudinal plasma mode forms a BIC and the transversal leaky modes. (c) Formation of the accidental BIC at $k_x\neq0$ due to the destructive interference of quasi-longitudinal plasma mode and quasi-transversal conventional mode.}
\label{fig:fig1}
\end{figure}

We consider a wire medium composed of finite-height thin metallic wires made of a perfect electric conductor (PEC) and arranged in a periodic array with a square unit cell [see Fig.~\ref{fig:fig1}(a)]. Such media are well-studied in literature~\cite{Belov2003Mar,Simovski2012Aug,Silveirinha2006Jun,Silveirinha2009Jan,Maslovski2002Oct}. They have the ability to manipulate electromagnetic waves in a way that traditional materials cannot.  Wire media can be designed to have a negative refractive index~\cite{silveirinha2010negative} or hyperbolic dispersion \cite{simovski2004low} and used for wave canalization~\cite{belov2005canalization}, cloaking~\cite{ktorza2015single}, and imaging with sub-wavelength resolution~\cite{ikonen2006experimental}. Within the EMA~\cite{Belov2003Mar}, the wire medium behaves as anisotropic 1D plasma with strong spatial dispersion that cannot be neglected even in the long-wave approximation when $a\ll \lambda$. Its dielectric tensor reads as~\cite{Belov2003Mar}:
\begin{align}
&\hat\varepsilon=
\mathrm{diag}[1,1, \varepsilon(\omega,k_z)], \label{Eq:eq-matrix} \\
&\varepsilon(\omega,k_z)=1-\frac{\Omega_p^2
}{\omega^2-c^2k_z^2},\label{Eq:eps}\\
  &\Omega_p^2 = \dfrac{{2\pi c^2/a^2}}{\ln\left({a/\pi d}\right) + 0.5275}. \label{Eq:plasma-freq}
\end{align}
Here, $d$ is the diameter of the PEC wires, $a$ is the lattice constant, and $\Omega_p$ is the plasma frequency of the wire medium.

%\begin{bmatrix}
%1 & 0 & 0 & \\
%0 & 1 & 0 & \\ 
%0 & 0 & \varepsilon(\omega,k_z) & 
%\end{bmatrix}

The spectrum of a bulk wire medium consists of two types of modes: (i) plasma waves, which are quasi-longitudinal and able to propagate only above the plasma frequency, and (ii) TEM modes, which can propagate at arbitrarily low frequencies. The dispersion of TEM modes does not depend on $k_x$ and $k_y$, and it is read as $k_z=\omega/c$~\cite{Simovski2012Aug}. Therefore, the isofrequency surface is completely flat, and energy propagates only along the wires. Finding the eigenmode spectrum in a slab of wire medium requires the introduction of additional boundary conditions~\cite{Silveirinha2006Jun,Silveirinha2008May}. They can be found from the assumptions of zero current at the end-faces of the wires. The derivations of the dispersion equation and the reflection coefficient from the slab of wire medium are derived in the Supplemental Material~\cite{supp}.   

\begin{figure}[t]
\centering
\includegraphics[width=0.99\linewidth]{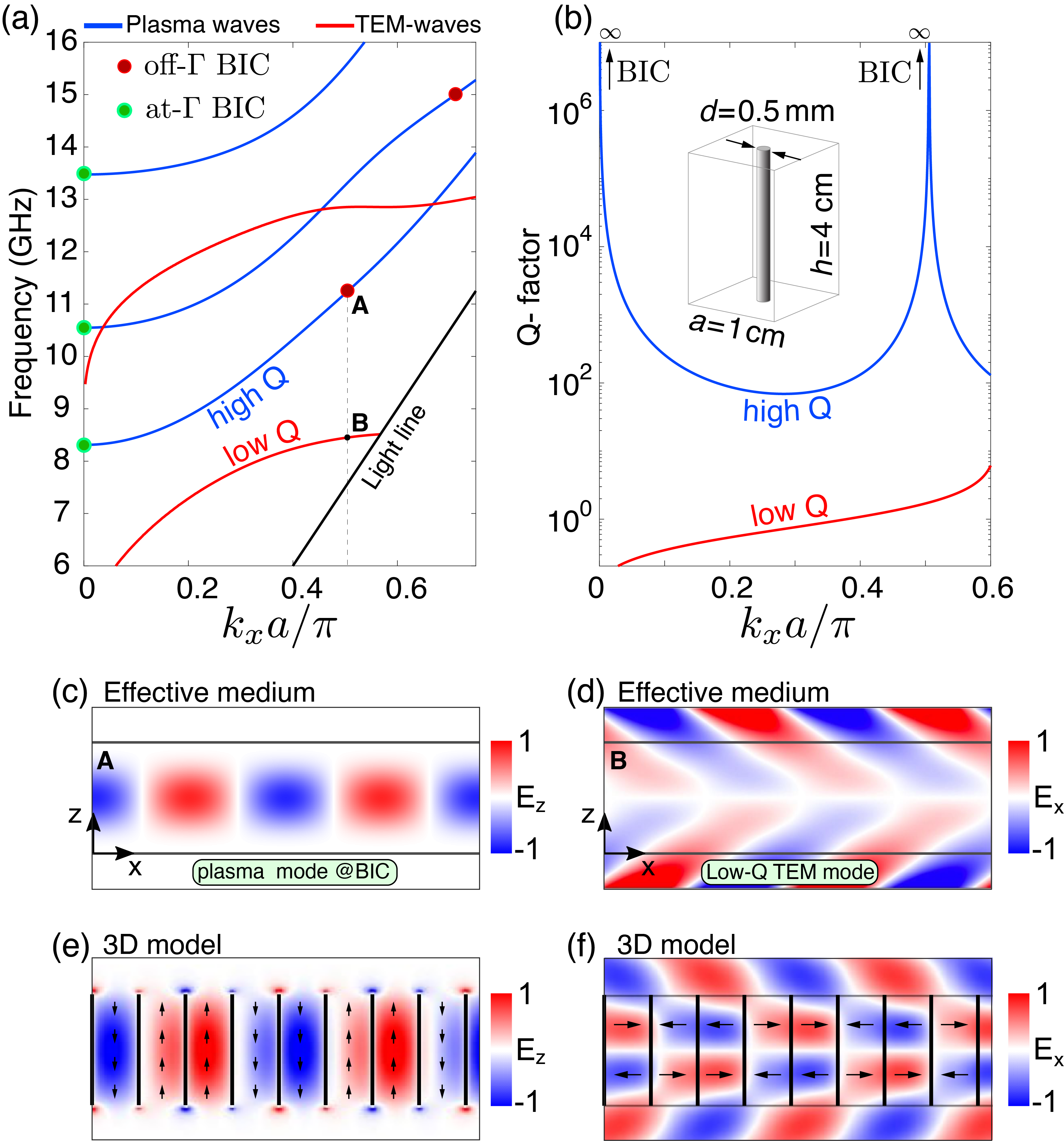}
\caption{(a) Dispersion and (b) Q-factor of the plasma and TEM-waves in a slab of wire medium calculated within EMA. The red and green dots mark accidental and symmetry-protected BICs, respectively. $E_z$-field distribution for at-$\Gamma$ BIC -- the plasma mode at $k_x=0$ calculated within EMA (c) and within the full 3D model (e).  $E_x$-field distribution for the low-Q TEM mode calculated within EMA (d) and within the full 3D model (f). Parameters of the wire medium: $a=10$~mm, $h=40$~mm, $d=0.5$~mm.} 
\label{fig:fig2}
\end{figure}

A slab of a wire medium naturally supports two types of BICs that can be described within EMA without taking into account a periodic structure. 
At $k_x=k_y=0$, the plasma modes are pure longitudinal ($\mathbf{E}\|\mathbf{k}$) and the average on unit cell magnetic field tends to zero $\mathbf{H}=0$. Such a mode cannot couple to the transversal waves in the surrounding space due to the polarization mismatch, and it turns into a symmetry-protected BIC [Fig.~\ref{fig:fig1}(b)]. This BIC can also be interpreted as the trapping of longitudinal plasma waves in an epsilon-near-zero medium~\cite{Demetriadou2008,Sakhno2021,Balafendiev2022}. Similar BICs appear in a slab of elastic material when the transverse (shear) acoustic wave does not couple to the longitudinal pressure waves in the surrounding fluid~\cite{Quotane2018Jan,Mizuno2019Feb,Deriy2022Feb}. An analogous mechanism of BIC formation was recently studied in double-net metamaterials~\cite{Wang2023Sep}.

At $k_x \ne 0$ or $k_y \ne 0$, both plasma and TEM waves are no longer independent. They mix, forming leaky TM modes, which can be called quasi-plasma and leaky quasi-TEM modes depending on the dominant component. This mixture can be interpreted as interaction via the radiation continuum as shown in Fig.~\ref{fig:fig1}(c), but formally, it is governed by the additional boundary conditions~\cite{Silveirinha2006Jun,Silveirinha2008May}. At certain values of $k_x$, the system can support the accidental BICs. They are formed when the radiative losses of the leaky mode are suppressed due to the destructive interference of plasma and TEM waves composing the mode (see the Supplemental Material for details~\cite{supp}). A similar mechanism of destructive interference between two modes can result in the appearance of accidental BICs in the photonic crystal slabs~\cite{Dong2021}.

Figure~\ref{fig:fig2}(a) shows the eigenmode spectrum for the wire medium slab calculated within EMA [see Eq.~(S1)]. The dimension axes are introduced for reference. They correspond to the following parameters of a wire medium: $a=10$~mm; diameter of wires $d=0.5$~mm; length of wires $h=40$~mm [see Fig.~\ref{fig:fig2}(b)]. Such parameters give plasma frequency $\Omega_p/(2\pi)=7.76$~GHz.
The red curves correspond to the quasi-TEM modes, while the blue curves correspond to the quasi-plasma modes discussed above. The divergence of the radiative Q-factor for $k_x=0$ and $k_x\neq 0$ shown in Fig.~\ref{fig:fig2}(b) manifests the presence of symmetry-protected and accidental BICs. The case of lossy wires is considered in the Supplemental Material~\cite{supp}. All accidental and symmetry-protected BICs are marked in Fig.~\ref{fig:fig2}(a) by red and green dots, respectively. It is worth mentioning that the accidental BICs in a wire medium appear as a result of destructive interference of bulk TEM and plasma modes existing in a wire medium slub in contrast to the majority of photonic systems where BICs appeared in the strong mode coupling regime~\cite{Koshelev2019Jun}. It means that there is no avoid-crossing phenomenon because the imaginary parts of energy of quasi-TEM and quasi-plasma modes are significantly different (see Supplemental Material). A similar behavior, for example, is observed in integrated Gires-Tournois interferometers where the interference between high-Q and low-Q modes results in the formation of BIC without avoiding crossing~\cite{bykov_bound_2020}. 

\begin{figure}[t] 
\centering
\includegraphics[width=0.99\linewidth]{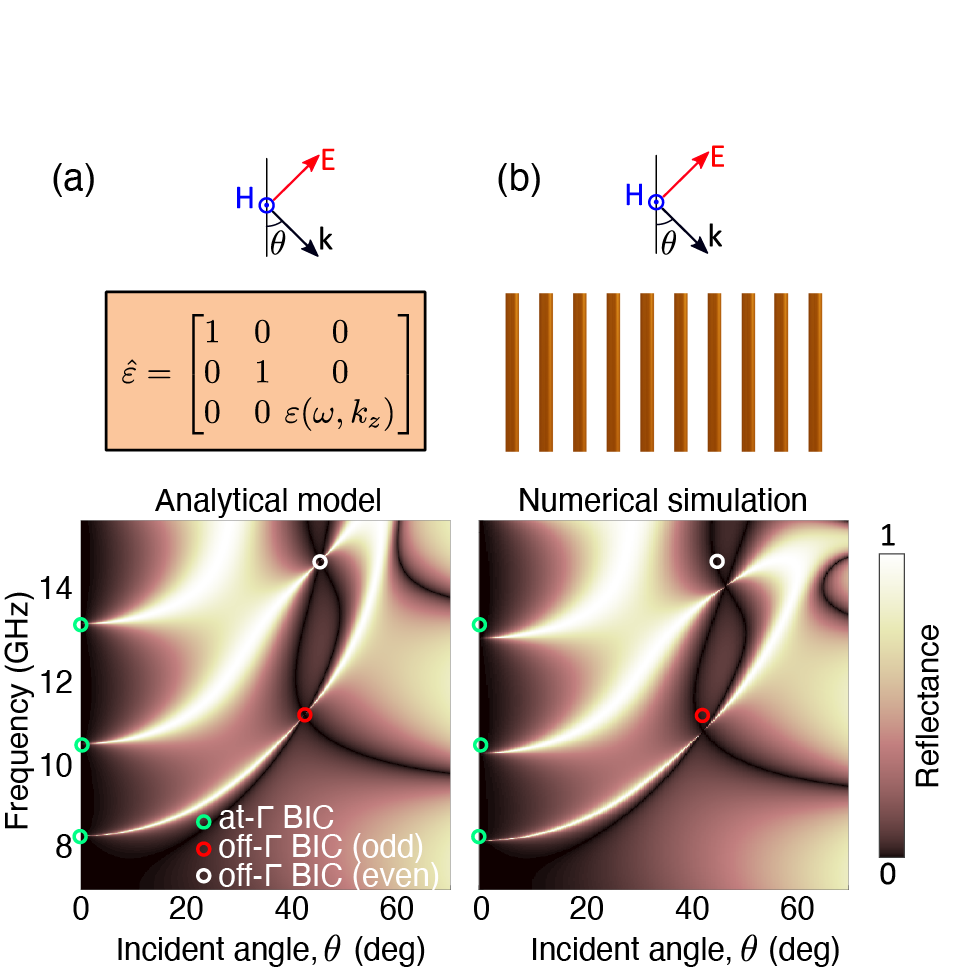} %0.99\linewidth
\caption{Map of reflection spectra calculated for various angles of incidence calculated analytically within effective medium approximation (a) and numerically with CST (b).} 
\label{fig:fig3}
\end{figure}
 
The full wave 3D model and EMA match well and give the agreed values of the spectral and angular positions of BICs. A more detailed comparison between EMA and 3D simulation is provided in the Supplemental Material~\cite{supp}.   However, the field distributions predicted from these two approaches are similar for the low-Q TEM modes [see Figs.~\ref{fig:fig2}(d) and \ref{fig:fig2}(f)] but substantially distinct for BIC [see Figs.~\ref{fig:fig2}(c) and \ref{fig:fig2}(e)]. In the full wave 3D model [see Figs.~\ref{fig:fig2}(e)], BIC does not have far-fields but has evanescent fields corresponding to the closed diffraction channels. Within the EMA [see Figs.~\ref{fig:fig2}(c)], BIC does not have both near- and far-fields. Thus, they become completely localized inside the slab. This is very unusual, as even the waveguide modes have evanescent fields. The vanishing electric field outside the slab can be used as a feature for finding the angular and spectral positions of BICs (see the Supplemental Material~\cite{supp}). 

The frequencies of the symmetry-protected BICs correspond to the Fabry-Perot resonance of the plasma wave across the slab:
\begin{align}
&\omega_m^2=(\pi m c/ h)^2 + \Omega_p^2; & & k_z=0.
\label{eq:BIC-at-Gamma}
\end{align}
Here $m$ is an integer. The frequencies of the accidental BICs correspond to the Fabry-Perot resonances of TEM mode, meanwhile, $k_z$ is defined from the dispersion equation~\cite{supp}:
\begin{align}
& \omega_n^2=(n \pi c/ h)^2;\nonumber \\ 
&k^2_z=(\omega_n^2-\Omega_p^2)/c^2-(m \pi c/ h)^2.
\label{eq:BIC-off-Gamma}
\end{align}
Here $n$ and $m$ are integers of the same parity (odd/even). 

% \begin{align}
% &\omega_n^2=(\pi m c/ h)^2 + \Omega_p^2; & & k_x=0.
% \label{eq:BIC-at-Gamma}
% \end{align}
% For accidental BICs:
% \begin{align}
% & \omega_n^2=(2\pi  c/ h)^2; & & c^2k^2_z=(\pi /h)^2(n^2-m^2)-\Omega_p^2,
% \label{eq:BIC-off-Gamma}
% \end{align}
% where $n$ and $m$ are integers.

%The accidental BICs in a wire medium appear in a weak coupling regime without characteristic avoid crossing in contrast to the majority of photonic systems where BICs appeared in the strong mode coupling regime~\cite{1}. 

%between plasma and TEM modes but not in a strong one as it usually happens in many photonic systems~\cite{1}. Indeed, one can see in Fig.~\ref{fig:fig2} that the real parts of the complex eigenfrequencies for the plasma and TEM modes cross each other while the imaginary ones repulsed. This can be explained by the large radiative losses for TEM mode. 

Figure~\ref{fig:fig3} shows the reflection spectra from the slab of wire medium calculated analytically within EMA [panel~(a)] and using the full-wave numerical simulation [panel~(b)]. The parameters of the medium are shown in the inset of Fig.~\ref{fig:fig2}(b). The incident waves are assumed to be p-polarized. Both methods are well agreed. Both symmetry-protected and accidental BICs manifest themselves as infinitely narrow peaks disappearing exactly at the BIC points. The circles in Figs.~\ref{fig:fig3}(a) and \ref{fig:fig3}(b) indicate the spectral and angular positions of symmetry-protected and accidental BICs calculated with Eqs.~\eqref{eq:BIC-at-Gamma} and \eqref{eq:BIC-off-Gamma}. The slight deviations are explained by the framework of EMA. Thus, for thinner and longer wires, the deviations are smaller.

\begin{figure}[t]
\centering
\includegraphics[width=0.95\linewidth]{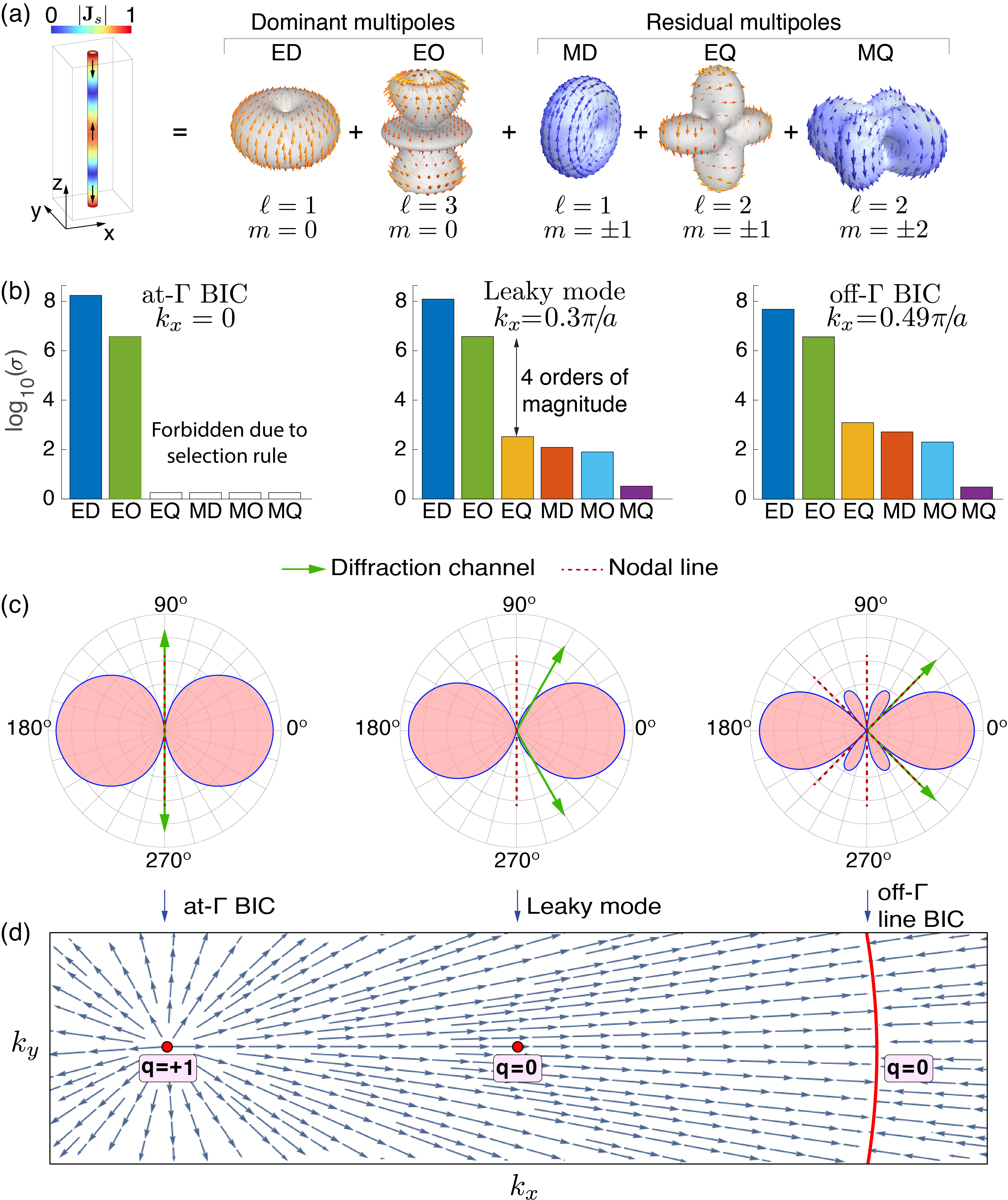}
\caption{(a) Characteristic distribution of surface current in a unit cell of a wire medium slab. Multipole series for plasma mode's near- and far-fields in the wire medium. (b) Multipolar decomposition of the plasma wave for different values propagation constant: symmetry-protected BIC ($k_x=0$); leaky mode regime ($k_x=0.3\pi/a$); accidental BIC ($k_x=0.49\pi/a$).  (c) Directivity diagrams of the resulting multipole for the same values of $k_x$ plotted in a logarithmic scale. The solid green arrows show the direction of open diffraction channels. The dashed red lines show the nodal lines for the multipole. (d) The far-field polarization map accounting for the contribution of ED and EO.} 
\label{fig:fig4}
\end{figure}

BIC, being a polarization vortex in the far field, can be characterized by {\it topological charge} defined as the winding number of the polarization vortex, which shows the number of counterclockwise rotations the electric field makes along the path in the k-space enclosed the BIC if going the counterclockwise direction~\cite{Zhen2014Dec,Hsu2013Jul}. Therefore, BIC is robust against variations of any parameters preserving the symmetry of the system~\cite{Bulgakov2017Dec,Liu2019Sep,Jin2019Oct}. On the other hand, a BIC appears when the nodal line of the directivity diagram of the unit cell coincides with the direction of the open diffraction channel~\cite{sadrieva2019multipolar,Chen2019Apr}. The directivity diagram and topological charge of BIC can be well described by a few dominant multipoles of the mode which can be predicted from the group theory without direct calculations~\cite{Zhen2014Dec,sadrieva2019multipolar,Gladyshev2022Jun}. To gain deeper insight into the topological nature of BICs in a wire medium, we analyze their multipolar composition using the group theory~\cite{sadrieva2019multipolar}.

Figure~\ref{fig:fig4}(a) shows the multipolar composition for the even plasma-like mode [$\mathbf{E}(-z)=\mathbf{E}(z)$] for the Bloch wavenumber along the $\Gamma \text{X}$ direction. The multipolar harmonics are proportional to $\sin m\varphi$ or $\cos m\varphi$, where $\varphi$ is the azimuthal angle and $m$ is an integer.  Because the unit cell is a thin wire ($d\ll a$ and $d\ll h$), the main contribution is given only by the electrical multipoles with the rotation symmetry ($m=0$), namely, vertical electric dipole, linear vertical electric octupole, etc. Figure~\ref{fig:fig4}(b) shows the amplitudes of the multipolar decomposition of the plasma-like mode for three different Bloch wavenumbers. The amplitudes are calculated by integrating the currents over the surface of the PEC wire~\cite{supp}. One can see that the multipoles without rotating symmetry ($m\neq0$) are suppressed at least by four orders of magnitude. This suppression ratio can be even higher for thinner wires. Figure~\ref{fig:fig4}(c) shows the numerically calculated directivity diagrams for the currents in the unit cell corresponding to at-$\Gamma$ BIC, leaky mode, and off-$\Gamma$ BIC. The Bloch wavenumbers are taken the same as in Fig.~\ref{fig:fig4}(b).  The calculations demonstrate that both accidental and off-$\Gamma$ BICs are formed when the nodal lines of the directivity diagram (dashed red lines) coincide with the direction of the open diffraction channel (green arrows). It is worth mentioning that the directivity diagram remains symmetric with respect to the vertical axis even for non-zero Bloch wavenumbers, indicating that the higher-order multipoles are suppressed.

BICs and the leaky states at the neighbor Bloch wavenumbers form a polarization vortex with a singularity at its center. Thus, BIC can be associated with polarization singularity carrying a topological charge which is usually defined as a winding number of the vortex~\cite{Zhen2014Dec}. The rigorous way for calculation of the topological charge $q$ of a BIC includes the finding of the far-field for the leaky states around the BIC followed by integration:
\begin{equation}
q=\frac{1}{2\pi}\oint_C\nabla_{\mathbf{k}}\phi(\mathbf{k}) \mathrm{d}\mathbf{k}, \quad q \in \mathbb{Z}. 
\label{eq:top_charge}
\end{equation}

Here  $\phi(\mathbf{k}) = \text{arg}[E_x(\mathbf{k}) + iE_y(\mathbf{k})]$, $E_x$ and $E_y$ the complex amplitudes of the radiated plane waves, and $C$ is a simple counterclockwise oriented path enclosing the singular point. However, the far-field polarization structure and topological charge can be predicted using the multipolar decomposition of the polarization currents in the unit cell~\cite{sadrieva2019multipolar}. In combination with the group theory, this is a powerful tool for the analysis of the far-field polarization singularities of metasurfaces and single nanostructures~\cite{Gladyshev2022Jun,Gladyshev2020Aug,Chen2019Apr,Volkovskaya2020Sep}.   

\begin{figure}[t]
\centering
\includegraphics[width=0.99\linewidth]{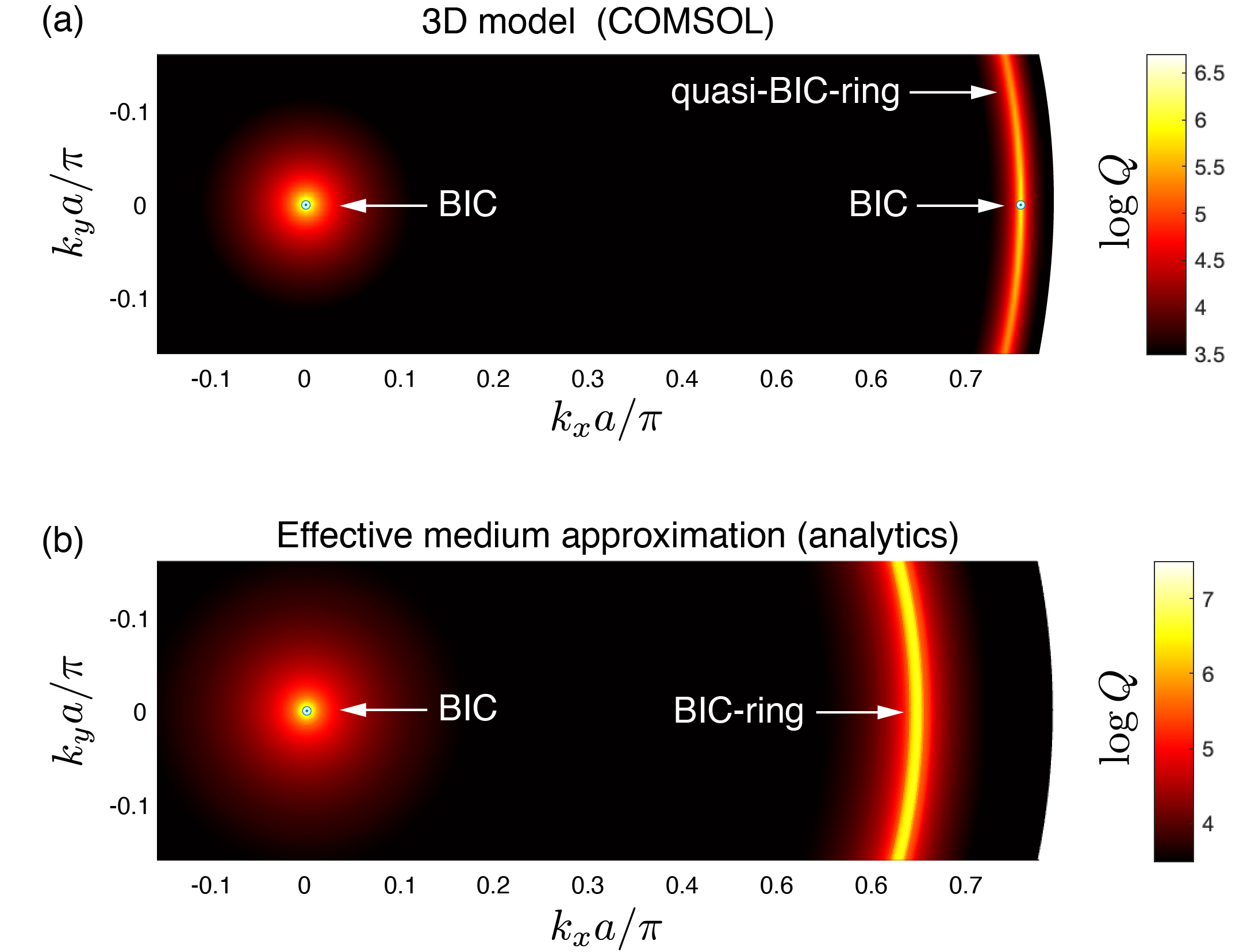}
\caption{The Q-factor maps of the plasma mode in a slab of wire medium as a function of $k_x$ and $k_y$ calculated (a) numerically in COMSOL Multiphysics and (b) analytically with EMA.} 
\label{fig:fig5}
\end{figure}

The EMA correctly predicts the frequency and angular position of symmetry-protected and accidental BICs [see Fig.~\ref{fig:fig3}] but does not completely describe their topological properties. The EMA neglects the internal structure of the wire medium and keeps only the multipoles with the rotational symmetry ($m=0$), i.e., all in-plane directions are assumed to be identical~\cite{Gorlach2015Aug,Petschulat2008Oct,Yaghjian2013Nov}. Therefore, the accidental BIC within EMA forms a ring in the k-space~\cite{Kostyukov2022Feb,Bulgakov2019Aug,Gladyshev2022Jun}.  Figure~\ref{fig:fig4}(d) shows the polarization map calculated with accounting for two dominant multipoles, ED and EO, whose amplitudes were calculated numerically in COMSOL~\cite{supp}. One can see that the topological charge of the symmetry-protected BIC is $q=1$,  while the accidental BIC forms a ring in the k-space for which topological charge $q$ is ill-defined with Eq.~\eqref{eq:top_charge}. The account for higher-order multipoles with $m\neq0$ transforms the ring of BICs to the ring of quasi-BIC with the genuine BICs only along the high-symmetry directions. However, it is difficult to track this transformation even numerically as the amplitudes of higher-order multipoles are substantially suppressed. To visualize this transformation, we increase the radius of the wires and consider a wire medium slab composed of the wires with diameter $d = 3$~mm and height $h = 4$~cm. The period of the structure is taken as previously $a = 1$~cm. Figure~\ref{fig:fig5} shows the Q-factor maps calculated numerically with COMSOL Multiphysics and analytically within EMA for the fundamental plasma-like mode. One can see that EMA gives a ring of BICs [Fig.~\ref{fig:fig5}(b)] while the full-wave numerical simulation shows that the ring of BICs is actually a ring of quasi-BICs and a genuine BIC exists only along the $\Gamma$X-direction [Fig.~\ref{fig:fig5}(a)]. The higher-order multipoles can be included in EMA that results in additional spatial dispersion, resulting in the dependence of the effective permittivity on the in-plane wavevector~\cite{capolino2017theory,chebykin_spatial-dispersion-induced_2015}. 

In summary, we have shown that the strong spatial dispersion in a slab of a wire medium composed of thin metallic wires results in the appearance of bound states in the continuum forming in an unusual way. The spatial dispersion makes possible propagation of longitudinal plasma-like waves. These waves form a symmetry-protected BIC in the $\Gamma$-point, as they polarization orthogonal to the transversal waves in the surrounding space. The accidental BICs are formed due to the destructive interference of plasma-like waves and TEM waves.
 Both types of BICs are well-described within effective medium approximation without the use of Bloch theorem and diffraction effects. We also show that the found BICs have a topological nature carrying an integer topological charge. An interesting feature of the revealed BICs is that they have no near-field within EMA, being completely localized inside the structure in contrast to the conventional BICs in periodic metasurfaces. This property of BICs can be prospective for avoiding cross-talks in compact integrated photonic circuits. The developed theory of BICs is general, and it can be equally applied to photonic, phononic, magnonic, and excitonic systems, enhancing the capabilities of acoustic, magnetic, and polaritonic devices as the spatial dispersion effects in such systems can play a crucial role~\cite{Basov2016Oct,Bossart2023May}.

\section*{Acknowledgement}
The authors acknowledge useful discussions with Dmitriy Maksimov, Kirill Koshelev, Yuri Kivshar, and Mihail Petrov. They also acknowledge Grigoriy Karsakov for technical support. Ivan Terekhov's work was financially supported by the ITMO Fellowship Program, the Federal program Priority 2030, the Russian Science Foundation, and the BASIS Foundation.  

\bibliography{bib.bib}

\end{document}

\title{\Large{Supplemental Material: \\ Bound states in the Continuum in a Wire Medium}}

\author{\firstname{E.}\,\surname{Koreshin}}
\affiliation{Qingdao Innovation and Development Base of Harbin Engineering University, Qingdao 266000, Shandong, China}
\affiliation{School of Physics and Engineering, ITMO University, Saint-Petersburg, Russia}
\author{\firstname{S.}\,\surname{Gladyshev}}
\affiliation{School of Physics and Engineering, ITMO University, Saint-Petersburg, Russia}
\author{\firstname{I.}\,\surname{Matchenya}}
\affiliation{School of Physics and Engineering, ITMO University, Saint-Petersburg, Russia}
\author{\firstname{R.}\,\surname{Balafendiev}}
\affiliation{School of Physics and Engineering, ITMO University, Saint-Petersburg, Russia}
\affiliation{Science Institute, University of Iceland, 107 Reykjavik, Iceland}
\author{\firstname{I.}\,\surname{Terekhov}}
\affiliation{School of Physics and Engineering, ITMO University, Saint-Petersburg, Russia}
\author{\firstname{P.}\,\surname{Belov}}
\affiliation{School of Physics and Engineering, ITMO University, Saint-Petersburg, Russia}
\author{\firstname{A.}\,\surname{Bogdanov}}
\email{a.bogdanov@metalab.ifmo.ru}
\affiliation{Qingdao Innovation and Development Base of Harbin Engineering University, Qingdao 266000, Shandong, China}
\affiliation{School of Physics and Engineering, ITMO University, Saint-Petersburg, Russia}

\date{\today}

\begin{abstract}

%In this Supplemental Material, we provide details of theoretical models for the description of symmetry-protected (at-$\Gamma$) and accidental (off-$\Gamma$) Bound States in the Continuum (BIC) in a wire medium described within effective medium approximation. In Sec.~SI, derive the reflection and transmission coefficients from the wire medium slab. In Sec.~SII, we derive the dispersion equation for the eigenmodes in the wire medium slab within effective medium approximation. In Secs.~SIII and SIV we provide the explicit expressions for spectral and angular positions of the accidental and symmetry-protected BICs. In Sec.~V, we analyze the behaviour of the solutions in the vicinity of BICs. In Sec.~VI, we descripencies between full-wave numerical simulations and effective medium model for a wire medium slab composed of non-thinner wires. 

In this Supplemental Material, we present a comprehensive analysis of theoretical models to describe symmetry-protected (at-$\Gamma$) and accidental (off-$\Gamma$) Bound States in the Continuum (BIC) in a wire medium, using the effective medium approximation. Section S.I details the derivation of reflection and transmission coefficients for a wire medium slab. Section S.II focuses on deriving the dispersion equation for the eigenmodes in the wire medium slab under the effective medium approximation. Sections S.III and S.IV provide explicit expressions for the spectral and angular positions of accidental and symmetry-protected BICs. Finally, In Section S.V, we examine the behavior of solutions near BICs in the scattering problem.

\end{abstract}

\keywords{}
\maketitle

\section{Reflection and Transmission coefficients from a slab of wire medium}

The permittivity tensor describing the wire medium depicted in Fig.~1 of the main text has the form~\cite{belov2006resolution}:
\begin{eqnarray}
&&\bar{\epsilon}_{ij}=\varepsilon(\omega,k_z) \bm n_i\bm n_j+(\delta_{ij}-\bm n_i\bm n_j)\,,\\
&&\varepsilon(\omega,k_z)=1-\frac{k_p^2}{k^2-k_z^2}\,,
\end{eqnarray}
where vector $\bm n$ directed along the $z$-axes $\mathbf{n}=(0,0,1)^T$,  $\delta_{ij}$ is the Kronecker delta symbol, $\mathbf{k}=(k_x,k_y,k_z)$ is the wave vector, $\bm k^2=\omega^2/c^2$, $c$ is the speed of light, $k_p=\Omega_p/c$,  $\Omega_p$ is the plasma frequency  \cite{belov2002}: 
\begin{eqnarray}
\Omega_p^2\approx\frac{2\pi c^2}{a^2}\left(\ln\frac{a}{\pi d}+ 0.5257 \right)^{-1}.
\end{eqnarray}
Here $d$ is the wire diameter,  $a$ is the lattice period.  

Let us consider the scattering problem of a plane wave with TEM polarization on the wire medium slab.  The slab is unbounded along the $x$ and $y$-directions and has the thickness $d$ in the $z$-direction.  Let the incident plane wave have the wave vector $\mathbf{k}=(k_x,0,k_z)$.  The magnetic field of the wave is parallel to the $y$-direction $\mathbf{H}=(0, H_y,0)$.  The incident wave excites two modes in the slab, namely, TM mode and transmission line TEM mode~\cite{belov2003}.  The wave vectors of the TM and TEM-modes are $\mathbf{q}^{\text{TM}}_{\pm}=(k_x,0,\pm\gamma)$, $\mathbf{q}^{\text{TEM}}_{\pm}=(k_x,0,\pm k)$, respectively \cite{belov2003},  where $\gamma=\sqrt{k^2-k_x^2-k_p^2}$. The magnetic field  $H_y(x,z)$ can be presented  in the form:
\begin{eqnarray}\label{ScattProb}
H_y(x,z)=H_0e^{-ik_x x}\left\{\begin{array}{ll}
e^{-ik_z (z+d/2)}+Re^{ik_z (z+h/2)},&z<-h/2,\\
A^{\text{TM}}_+ \cos(\gamma z)+B^{\text{TEM}}_+ \cos(k z)+A^{\text{TM}}_- \sin(\gamma z)+B^{\text{TEM}}_- \sin(k z),&|z|\le h/2,\\
Te^{-ik_z (z-h/2)},&z>h/2,
\end{array}\right.
\end{eqnarray} 
where $H_0$ is the magnetic field amplitude.  The reflection $R$,  transmission $T$, and $A^{\text{TM}}_{\pm}$,  $B^{\text{TEM}}_\pm$ can be found from the matching conditions at the slab boundaries.  It was shown that at the boundary of a wire medium, the magnetic field as well as its first and second derivatives are continuous at the slab boundaries \cite{Silveirinha2006}.  The matching conditions for magnetic field give \cite{belov2006resolution}:
\begin{eqnarray}\label{R}
&&R=1-\frac{ik_z(k_x^2+k_p^2)\cos\frac{\gamma h}{2}\cos\frac{kh}{2}}{F_+(k,k_x)}-\frac{ik_z(k_x^2+k_p^2)\sin\frac{\gamma h}{2}\sin\frac{kh}{2}}{F_-(k,k_x)}\,,\label{R}\\
&&T=\frac{ik_z(k_x^2+k_p^2)\cos\frac{\gamma h}{2}\cos\frac{k h}{2}}{F_+(k,k_x)}-\frac{ik_z(k_x^2+k_p^2)\sin\frac{\gamma h}{2}\sin\frac{k h}{2}}{F_-(k,k_x)}\,,\label{T}\\
&&A^{\text{TM}}_+=\frac{ik_zk_x^2\cos\frac{kh}{2}}{F_+}\,, \quad A^{\text{TM}}_-=-\frac{ik_z k_x^2\sin\frac{kh}{2}}{F_-},\label{Apm}\\
&&B^{\text{TEM}}_+=\frac{ik_zk_p^2\cos\frac{\gamma h}{2}}{F_+}\,, \quad B^{\text{TEM}}_-=-\frac{ik_zk_p^2\sin\frac{\gamma h}{2}}{F_-}\label{Bpm},
\end{eqnarray}
where 
\begin{eqnarray}
F_+(k,k_x)&=&i k_z(k_x^2+k_p^2)\cos\frac{\gamma h}{2}\cos\frac{k h}{2}-\gamma k_x^2\sin\frac{\gamma h}{2}\cos\frac{kh}{2}-k\,k_p^2\cos\frac{\gamma h}{2}\sin\frac{k h}{2}\,\label{F_p}\\
F_-(k,k_x)&=&i k_z(k_x^2+k_p^2)\sin\frac{\gamma h}{2}\sin\frac{k h}{2}+\gamma k_x^2\cos\frac{\gamma h}{2}\sin\frac{k h}{2}+k\,k_p^2\sin\frac{\gamma h}{2}\cos\frac{k h}{2}\, \label{F_m}.
\end{eqnarray}
The complex zeros of $F_+(k,k_x)$ and $F_-(k,k_x)$ correspond to the symmetric and antisymmetric modes of the wire medium slab. Therefore, the dispersion equations for them can be written as

\begin{eqnarray}
&\text{Symmetric modes: \ \ \ \ \ \ \ \ \ \ \ } &F_+(k,k_x)=0\label{F_pEq0}, \\
&\text{Anti-symmetric modes: \ \ \ \ \ }& F_-(k,k_x)=0\label{F_mEq0}.
\end{eqnarray} 

\section{Diespersion equation}

Let us explicitly derive the dispersion equation for the eigenmodes in a wire medium. For instance, let us consider a symmetric mode. Its magnetic field can be written as:
\begin{eqnarray}\label{Sol_Incorrect1}
H_y(z)=\left\{
\begin{array}{ll}
Ce^{-ik_z(z-h/2)},&z>h/2,\\
A \cos\gamma z+B \cos kz,&|z|\leq h/2,\\
Ce^{ik_z(z+h/2)},& z<-h/2.
\end{array}\right. 
\label{eq:m_field}
\end{eqnarray}
The magnetic field as well as its first and second derivatives are continuous at the slab boundaries \cite{Silveirinha2006}. The matching conditions lead to the following system of equations:
\begin{eqnarray}\label{matchCondLeaky}
\left(
\begin{array}{ccc}
\cos(\gamma h/2) & \cos(k h/2)&-1\\
\gamma\sin(\gamma h/2) & k\sin(kh/2)&-i k_z\\
\gamma^2\cos(\gamma h/2)& k^2\cos(kh/2)&-k_z^2
\end{array} 
\right)\left(
\begin{array}{c}
A\\
B\\
C
\end{array}\right)=
\left(
\begin{array}{c}
0\\
0\\
0
\end{array}\right).
\end{eqnarray}
The nontrivial solution of the equations exists when the determinant of the matrix in Eq.  \eqref{matchCondLeaky}
equals to zero. This condition leads to equation \eqref{F_pEq0}.  The same consideration for the anti-symmetric solution leads to Eq.~\eqref{F_mEq0}. The solutions of Eqs.~\eqref{F_pEq0}, \eqref{F_mEq0} are presented in Fig.~2 of the main text.   The solutions $\omega(k_x)$ of Eqs.  \eqref{F_pEq0} and \eqref{F_mEq0}  have real and imaginary parts.  The positions of the BICs are located on the curve corresponding to the solution $\omega(k_x)$ which has a small imaginary part.  When  $k=k_{\pm}$ and $\gamma=\gamma_{\pm}$ the imaginary part of $\omega$ equals to zero.  At such parameters $C=0$ we obtain BIC. The same consideration for the anti-symmetric solution leads to Eq.~\eqref{F_mEq0}.

%Let us find the dispersion equation for the eigenmodes in a wire medium. To find a symmetric leaky TEM mode we present the magnetic field in the form:
%\begin{eqnarray}\label{Sol_Incorrect1}
%H_z(x,z)=H_0e^{-ik_x x}\left\{
%\begin{array}{ll}
%Ce^{-ik_z(z-h/2)},&z>h/2,\\
%A \cos k z,&|z|\leq h/2,\\
%Ce^{ik_z(z+h/2)},&z<-h/2.
%\end{array}\right. 
%\end{eqnarray}
%The matching conditions at the slab boundaries lead to the following equations:
%\begin{eqnarray}
%A\cos \frac{k h}{2}&=&C,\label{First1}\\
%k A\sin\frac{k h}{2}&=&ik_z C,\label{First2}\\
%k^2A\cos \frac{k h}{2}&=&k_z^2C\label{First3}\,.
%\end{eqnarray}
%Using equation \eqref{First1},  \eqref{First3},  and taking into account the relation  $k^2=k_z^2+k_x^2$,  we obtain:
%$$ k_x^2\cos\frac{k h}{2}=0.$$
%For nonzero $k_x$ there is only trivial solution $A=C=0$,  since in the case $\cos\frac{k h}{2}=0$,  Eq.  \eqref{First2} can not be fulfilled.  Therefore there are no leaky solutions for the TEM mode.  One can obtain similar results for the plasma mode.  So,  there are no pure TEM or TEM leaky modes.  
%

\section{accidental  off-$\Gamma$ BIC}
The spectral and angular position of off-$\Gamma$-BICs can be found as real solutions of Eqs.~\eqref{F_p} and \eqref{F_m}. One can see, that Eq.~\eqref{F_p} when 
\begin{equation}
\cos\frac{\gamma h}{2}=\cos\frac{k h}{2}=0.
\end{equation}
The solutions of these equations can be written as
\begin{eqnarray}\label{k+}
k&=&k_{+}=\frac{\pi(2n+1)}{h} ,\,\,\gamma=\gamma_{+}=\frac{\pi(2m+1)}{h},\\
k_x&=&k_{x+} =\pm\sqrt{k_{+}^2-\gamma_{+}^2-k_p^2},
\end{eqnarray}
where $m$ and $n$ equal to $0,1,2,\ldots$. $k = \omega/c$ is the wave vector of free-space.  Since $k_x$ must be real,  we have
\begin{eqnarray}
k_{+}^2-\gamma_{+}^2\ge k_p^2.
\end{eqnarray}

For the antisymmetric off-$\Gamma$-BIC the solution is the following:
\begin{eqnarray}\label{k-}
k&=&k_{-}=\frac{2\pi n}{h},\,\,\gamma=\gamma_{-}=\frac{2\pi m}{h},\\
k_x&=&k_{x-} =\pm\sqrt{k_{-}^2-\gamma_{-}^2-k_p^2},
\end{eqnarray}
where  $m$ and $n$ are equal to $1,2,3,\ldots$,   $k_{-}^2-\gamma_{-}^2\ge k_p^2$.  

Let us explicitly demonstrate that the first and the second sets indeed correspond to the symmetric and the anti-symmetric off-$\Gamma$ BICs.  Outside the slab the solution of Maxwell's equations are the incoming and outgoing plane waves,  therefore to find the stationary localized in the slab solution decreasing at $x\to\pm\infty$ we must set the magnetic field outside the slab equal to zero. 
Therefore to find  the symmetric BIC we present the magnetic field in the form: 
\begin{eqnarray}\label{SymSol}
H_+^{\text{BIC}}(x,z)=H_0 e^{-ik_x x}\left\{
\begin{array}{lc}
0, & |z|>h/2,\\
A^{\text{TM}}_+ \cos(\gamma z)+B^{\text{TEM}}_+ \cos(k z), &|z|\le h/2,\\
\end{array}
\right.
\end{eqnarray} 
The matching conditions at slab boundaries give three equations
\begin{eqnarray}\label{matchCond}
\left(
\begin{array}{cc}
\cos(\gamma h/2) & \cos(k h/2)\\
\gamma\sin(\gamma h/2) & k\sin(kh/2)\\
\gamma^2\cos(\gamma h/2)& k^2\cos(kh/2)
\end{array} 
\right)\left(
\begin{array}{c}
A^{\text{TM}}_+\\
B^{\text{TEM}}_+
\end{array}\right)=
\left(
\begin{array}{c}
0\\
0\\
0
\end{array}\right).
\end{eqnarray}
The system of equations \eqref{matchCond} has a nontrivial solution when the rank of the matrix equals to unity.
This condition is fulfilled for $k=k_+$ and $\gamma=\gamma_+$, see Eqs. \eqref{k+}.  For such $k$ and $\gamma$ we obtain: 
\begin{eqnarray}
&&A^{\text{TM}}_+=\frac{2n+1}{2m+1}(-1)^{n+m+1}B^{\text{TEM}}_+\,.
\end{eqnarray}
For the anti-symmetric BICs, we search for the solution in the form
\begin{eqnarray}\label{ASymSol}
H_-^{\text{BIC}}(x,z)=H_0 e^{-ik_x x}\left\{
\begin{array}{cc}
0, & |z|>h/2,\\
A^{\text{TM}}_- \sin(\gamma z)+B^{\text{TEM}}_- \sin(k z), &|z|\le h/2,\\
\end{array}
\right.
\end{eqnarray}
The matching conditions give the second set of solutions for $k$ and $\gamma$:  $k=k_-$,  $\gamma=\gamma_-$,  see \eqref{k-}.  For anti-symmetric BICs we have 
\begin{eqnarray}
A^{\text{TM}}_-=\frac{n}{m}(-1)^{n+m+1}B^{\text{TEM}}_-\,.
\end{eqnarray}
So,  we have found the parameters of the symmetric and anti-symmetric BICs and demonstrate that the frequencies and wave vectors of the  off-$\Gamma$ BICs are solutions of Eqs.  \eqref{F_pEq0} and  \eqref{F_mEq0}.
Note that the off-$\Gamma$ BICs are the interference of the plasma-like mode and TEM-mode coexisting in the slab.

\section{Leaky modes}
Let us find the leaky TEM and plasma modes. To find symmetric leaky TEM mode we present the magnetic field in the form:
\begin{eqnarray}\label{Sol_Incorrect1}
H_z(x,z)=H_0e^{-ik_x x}\left\{
\begin{array}{ll}
Ce^{-ik_z(z-h/2)},&z>h/2,\\
A \cos k z,&|z|\leq h/2,\\
Ce^{ik_z(z+h/2)},&z<-h/2.
\end{array}\right. 
\end{eqnarray}
The matching conditions at the slab boundaries lead to the following equations:
\begin{eqnarray}
A\cos \frac{k h}{2}&=&C,\label{First1}\\
k A\sin\frac{k h}{2}&=&ik_z C,\label{First2}\\
k^2A\cos \frac{k h}{2}&=&k_z^2C\label{First3}\,.
\end{eqnarray}
Using equation \eqref{First1},  \eqref{First3},  and taking into account the relation  $k^2=k_z^2+k_x^2$,  we obtain:
$$ k_x^2\cos\frac{k h}{2}=0.$$
For nonzero $k_x$ there is only trivial solution $A=C=0$,  since in the case $\cos\frac{k h}{2}=0$  Eq.  \eqref{First2} can not be fulfilled.  Therefore there is not leaky solutions for the TEM mode.  One can obtain similar result for the plasma mode.  So,  there is no pure TEM or TEM leaky modes.  To find the symmetric leaky mode we present the magnetic field in the form:
\begin{eqnarray}\label{Sol_Incorrect1}
H_y(z)=\left\{
\begin{array}{ll}
Ce^{-ik_z(z-h/2)},&z>h/2,\\
A \cos\gamma z+B \cos kz,&|z|\leq h/2,\\
Ce^{ik_z(z+h/2)},& z<-h/2.
\end{array}\right. 
\end{eqnarray}
The matching conditions lead to the system
\begin{eqnarray}\label{matchCondLeaky}
\left(
\begin{array}{ccc}
\cos(\gamma h/2) & \cos(k h/2)&-1\\
\gamma\sin(\gamma h/2) & k\sin(kh/2)&-i k_z\\
\gamma^2\cos(\gamma h/2)& k^2\cos(kh/2)&-k_z^2
\end{array} 
\right)\left(
\begin{array}{c}
A\\
B\\
C
\end{array}\right)=
\left(
\begin{array}{c}
0\\
0\\
0
\end{array}\right).
\end{eqnarray}
The nontrivial solution of the equations exists when the determinant of the matrix in Eq.  \eqref{matchCondLeaky}
equals to zero. This condition leads to equation \eqref{F_pEq0}.  The same consideration for the anti-symmetric solution leads to Eq.~\eqref{F_mEq0}. The solutions of Eqs.~\eqref{F_pEq0}, \eqref{F_mEq0} are presented in Fig.~2 of the main text.   The solutions $\omega(k_x)$ of Eqs.  \eqref{F_pEq0} and \eqref{F_mEq0}  have real and imaginary parts.  The positions of the BICs are located on the  on the blue curve corresponding to the solution $\omega(k_x)$ which has the small imaginary part.  When  $k=k_{\pm}$ and $\gamma=\gamma_{\pm}$ the imaginary part of $\omega$ equals to zero.  At such parameters $C=0$ and we obtain BIC. The red curve associated with quasi-TEM mode obtain large imaginary part for any $k_x$. The same consideration for the anti-symmetric solution leads to Eq.~\eqref{F_mEq0}.

\section{Symmetry-protected at $\Gamma$ BIC}
At the $\Gamma$-point ($k_x=k_y=0$),  therefore the equations for the BICs parameters are
\begin{equation}
F_+(k,0)=0,\quad F_-(k,0)=0.
\end{equation}
These equations lead to the following condition 
\begin{eqnarray}
\cos\frac{{\gamma}h}{2}\sin\frac{{\gamma}h}{2}=0.
\end{eqnarray}
The substitution $\cos\frac{{\gamma}h}{2}=0$ or $\sin\frac{{\gamma}h}{2}=0$ to the matching conditions for the magnetic field leads to the existence only the trivial solutions: $A^{\text{TM}}_{\pm}=B^{\text{TEM}}_{\pm}=0$.  The reason for that is in the case $k_x=0$ the electric and magnetic fields are orthogonal to the wire medium,  therefore there is no interaction between such a wave and the slab.  Therefore the at-$\Gamma$ point BICs can not have the structure of the magnetic field in the form \eqref{SymSol} or \eqref{ASymSol}.  However,  in the case $k_x=0$ there are longitudinal  Langmuir waves,  i.e. the pure electric waves with the electric field $\mathbf{E}$ parallel to the wires.  It is well known that for such waves the displacement field obeys the equation \cite{LLt10}
\begin{eqnarray}
\bm k\cdot\bm D=D_z=0.
\end{eqnarray}   
Since $D_\alpha=\varepsilon_{\alpha\beta}E_\beta$, and $E_z\ne 0$, we obtain that 
$$\varepsilon_{zz}=0=1-\frac{k_p^2}{k^2-k_z^2}.$$
Therefore the dispersion relation for Langmuir waves has form:
\begin{eqnarray}
k^2=k_p^2+k_z^2.
\end{eqnarray}
The boundary conditions for $E_z$ are
\begin{eqnarray}
&&\int\limits_{-h/2}^{h/2}dz\frac{d E_z}{dz}=E_z(h/2)-E_z(-h/2)=0,\\
&&E_z(\pm h/2)=0.
\end{eqnarray}
The first condition means that the total charge in the slab equals to zero. The second condition means that the current at the slab boundaries equals to zero ($j_z=\sigma E_z|_{z=\pm h/2}=0$). The boundary conditions lead to the quantization of $k_z$:
\begin{eqnarray}
k_z&=&\frac{2\pi}{h}\left(n+\frac{1}{2}\right), \\
k_z&=&\frac{2\pi m}{h}, 
\end{eqnarray}
where $n=0,1,2,\ldots$,  $m=1,2,3,\ldots$.  The first and the second solutions correspond to the symmetric and anti-symmetric at-$\Gamma$ point BICs,  respectively.

\section{Scattering problem in the vicinity of BIC} 
Let us consider in detail the scattering problem \eqref{ScattProb}.  Without loss of generality we restrict the consideration only the case of the symmetric BICs.  In the vicinity of the symmetric BIC  we substitute  $k$ and $\gamma$ in the form $k=k_++\delta k$,  $\gamma=\gamma_++\delta\gamma$ to \eqref{Apm} and \eqref{Bpm}, expand obtained expressions over small $|\delta k|$ and $|\delta\gamma|$,  and obtain in the leading order in $\delta k$ and $\delta\gamma$:
\begin{eqnarray}
A^{\text{TM}}_+&\approx &\frac{i(-1)^{m+1}k_{z}k_{y+}^2\delta k}{\gamma_+k_{y+}^2\delta k+k_+k_p^2\delta\gamma}\,,\\
B^{\text{TEM}}_+&\approx &\frac{i(-1)^{n+1}k_zk_p^2\,\delta\gamma}{\gamma_+k_{y+}^2\delta k+k_+k_p^2\delta\gamma}\,,\\
A^{\text{TM}}_-&\approx &-\frac{k_x^2}{k_x^2+k_p^2}\sin\frac{\gamma_+h}{2}\,,\\
B^{\text{TEM}}_+&\approx &-\frac{k_p^2}{k_x^2+k_p^2}\sin\frac{k_+h}{2}\,.
\end{eqnarray}
One can check that the values of coefficients $A^{\text{TM}}_+$ and $B^{\text{TEM}}_+$ at $\delta k=0$,  $\delta\gamma=0$ depend on the order of the limits $\delta k\to 0$ and $\delta \gamma\to 0$.  Indeed,  if we calculate first the limit $\delta k\to 0$,  then $\delta\gamma\to 0$
we obtain:
\begin{eqnarray}
A^{\text{TM}}_+=0,\,\,B^{\text{TEM}}_+= \frac{i(-1)^{n+1}k_{z+}\,}{k_+},
\end{eqnarray} 
but in the apposite case: $\delta\gamma\to 0$,  then $\delta k\to 0$ we find different result: 
\begin{eqnarray}
A^{\text{TM}}_+=\frac{i(-1)^{m+1}k_{z+}}{\gamma_+},\,\,B^{\text{TEM}}_+=0.
\end{eqnarray}
The reason of the dependence of the limits on the path is the spectrum degeneracy at the point $k=k_+$,  $\gamma=\gamma_+$. At the point we have the scattering problem solution and the BIC.

% \section{Ring of BICs}
%
%\begin{figure}
%\begin{center}
%\includegraphics[width=0.6\linewidth]{Supp-Fig-v2.png}
%\caption{The Q-factor maps of the plasma mode in a slab of wire medium as a function of $k_x$ and $k_y$ calculated (a) numerically (COMSOL) and (b) analytically (EMA).} 
%\label{fig:figS1}
%\end{center}
%\end{figure}
%
%
%The effective medium approximation (EMA) well describes the wire medium composed of thin wires. The discrepancies between EMA and the results of full-wave numerical simulations become more evident with the increase in the diameter of the wire. To demonstrate this we consider a wire medium slab composed of the wires with diameter $d = 3$~mm and height $h = 4$~cm. The period of the structure $a = 1$~cm. 
%
%Figure~\ref{fig:figS1} shows the Q-factor maps calculated numerically with COMSOL Multiphysics and analytically within EMA for the fundamental plasma-like mode. One can see that both models predict the symmetry-protected and accidental BICs. However, the EMA predicts a ring of BICs while the full-wave numerical simulation shows that the ring of BICs is actually a ring of quasi-BICs and a genuine BIC exists only along the $\Gamma$X-direction. This can be explained by the fact the EMA neglects the higher-order multipoles with $m>0$.    